\documentclass[aps,prb,twocolumn,showkeys]{revtex4}
\usepackage[tbtags]{amsmath}
\usepackage{graphics,graphicx,epsfig,amssymb,
xcolor,hyperref,latexsym,float,bm,bbm,times,adjustbox,braket,verbatim}
\usepackage[english]{babel}
\usepackage[utf8]{inputenc}
\newcommand{\icol}[1]{
    \left(\begin{smallmatrix}#1\end{smallmatrix}\right)%
}
\newcommand{\irow}[1]{
    \begin{smallmatrix}\left(\,#1\,\right)\end{smallmatrix}%
}

\begin{document}

\title{Thermal transitions of the modulated superfluid for
spin-orbit coupled correlated bosons\\ in an optical lattice}

\author{Arijit Dutta$^{1}$, Abhishek Joshi$^{1}$,
K. Sengupta$^2$ and Pinaki Majumdar$^{1}$}

\affiliation{
$^1$ Harish-Chandra Research Institute,
HBNI, Chhatnag Road, Jhunsi, Allahabad 211019, India\\
$^2$ School of Physical Sciences, Indian Association for the
Cultivation of Science, Jadavpur, Kolkata-700032, India.}
\date{\today}


\begin{abstract}
We investigate the thermal physics of a Bose-Hubbard model with Rashba
spin-orbit coupling starting from a strong coupling mean-field ground
state. The essential role of the spin-orbit coupling $\left(\gamma\right)$
is to promote condensation of the bosons at a finite wavevector $\bm{k}_{0}$.
We find that the bosons display either homogeneous or phase-twisted or
orbital ordered superfluid phases, depending on $\gamma$ and the inter-species
interaction strength $\lambda$. We show that an increase of $\gamma$ leads
to suppression of the critical interaction $U_c$ for the superfluid to Mott
insulator transition in the ground state, and a reduction of the $T_c$ for
superfluid to Bose-liquid transition at a fixed interaction.  We capture
the thermal broadening in the momentum distribution function, and the real
space profiles of the thermally disordered magnetic textures, including
their homogenization for $T \gtrsim T_{c}$. We provide a Landau theory
based description of the ground state phase boundaries and thermal transition
scales, and discuss experiments which can test our theory.
\end{abstract}

\keywords{spin-orbit coupling, Bose-Hubbard model}

\maketitle

\section{Introduction}

The physics of strong correlation in ultracold atom systems has been
a subject of intense theoretical and experimental research in the
recent past \cite{rev1,greiner1,expt1,expt2,bht1,bht2}. The initial studies
in this field concentrated on single boson species. This choice is
motivated by the experimental ease of realizing the superfluid (SF)
and Mott insulating (MI) states of these bosons. Indeed, the first
experimental study of SF-MI quantum phase transition used $^{87}{\rm
Rb}$ bosons in their $F=1$ state \cite{greiner1}. More recently,
there have been concrete proposals to realize artificial Abelian
gauge fields for such bosons \cite{gauge1, gauge2}. The phase
diagram of strongly correlated bosons in the presence of such gauge
fields have also been investigated \cite{abeth1,abeth2} and reveal
a rich structure.

Several recent cold atomic experiments
tune Raman processes to create artificial spin-orbit couplings
in multicomponent Bose systems
\cite{spielman2011,ketterle2016,ketterle2017}. Most of these
experimental procedures produce an equal mixture of Rashbha and
Dresselhaus coupling, which leads to an effective Abelian gauge field
for the bosons. However, there have been concrete proposals to
experimentally realize purely Rashba type spin-orbit
coupling \cite{spielman-review}. This
is equivalent to a non-Abelian
gauge-field for two component bosons.

The ground state phase diagram
of such systems have been theoretically studied \cite{iskin,
nandini-prl, hofstetter, kush1, saptarshi1}. These studies employed
several theoretical techniques such as mean field
theories\cite{iskin}, simulated annealing of effective quantum spin
models\cite{nandini-prl}, real space bosonic dynamical mean field
theory (BDMFT) \cite{hofstetter}, and strong coupling expansion
\cite{kush1,saptarshi1}. They have unearthed a rich ground
state phase diagram for these systems. Some of the unconventional
phases found include those with long range magnetic order in
the Mott ground state \cite{nandini-prl} and the possibility of a
boson condensate at finite momentum \cite{kush1,saptarshi1}.  Such
studies have also been supplemented by their weak-coupling
counterparts in the continuum where there is no Mott transition.
The weakly interacting condensates have been
studied using the Bogoliubov-Hartree-Fock approximation
\cite{baym2014}.

In spite of several studies on the ground state, only limited
theoretical work exists on the thermal phases of spin-orbit coupled
systems. For Abelian systems with equal mixture of Rashba and
Dresselhaus coupling, Ref.\ \onlinecite{hickey} derives an effective
$t-J$ model for the bosons and studies the thermal phases of this
effective model. The study reveals a stripe superfluid order at low
temperature and a two step melting upon increasing temperature,
leading first to a striped normal phase of the bosons and then to a
homogeneous state. Similar studies were carried out for two
component fermions in optical lattices \cite{ref51}. However, to the
best of our knowledge, the thermal phases of Bose-Einstein
condensates (BECs) in the presence of Rashba spin-orbit coupling
have not been studied before. This is particularly pertinent since
an equal mixture of Rashba and Dresselhaus terms breaks the
four-fold rotation symmetry of the lattice, while the Rashba
spin-orbit term keeps it intact. This leads to the possibility of
superfluid phases with lower symmetry than that of the lattice.

In this work, we study the thermal phases of a two-orbital
Bose-Hubbard model in the presence of a Rashba spin-orbit coupling.
Our study thus involves bosons in the presence of an effective
non-Abelian gauge field. In what follows, we use an auxiliary field
decomposition of the kinetic energy followed by a `classical'
approximation to the auxiliary field. We then carry out a
Monte-Carlo study of the resulting model, sampling the auxiliary
field configurations. The method has been used in the past for the
single species Bose-Hubbard model \cite{joshi-thermal}. It retains
the key low energy thermal fluctuations and yields accurate thermal
transition scales.

We start by deriving an effective Hamiltonian whose
mean field ground state coincides, in the main, with
earlier results \cite{nandini-prl}.
Our results on this problem are the following:
(i)~We
find that the ground state is either a Mott insulator, or
a superfluid with condensation either
at a single wavevector $({\bf k}_0)$
or two wavevectors $(\pm {\bf k}_0)$.
The $\pm {\bf k}_0$  condensate constitutes a
orbital density wave, while the finite ${\bf k}_0$
condensate is a phase twisted superfluid \cite{saptarshi1}.
(ii)~The superfluid has associated `magnetic'
textures - related to the spatially varying orbital
occupancy. (iii)~Increasing temperature leads to the
simultaneous loss of superfluidity and order in the magnetic
textures. We establish the $T_c$ scale for varying Hubbard
interaction, interspecies coupling and spin-orbit interaction
using our Monte Carlo scheme \cite{joshi-thermal}.
(iv)~The momentum distribution function, $n_{\bm{k}}$,
evolves from its
`low symmetry' character at low temperature
to four-fold symmetry as $T \rightarrow T_c$,
providing a detectable thermal signature
of Rashba coupling.
Finally, (v)~we
construct an effective Landau theory which provides some
analytic understanding of the thermal scales, and discuss
experiments which can test our theory.

The plan of the rest of this work is as follows. In Sec.\
\ref{modelmethod}, we introduce the Bose-Hubbard Hamiltonian in the
presence of Rashba spin-orbit coupling and describe the method used
for our calculation. This is followed by Sec.\ \ref{gs}, where we
study the ground state phase diagram. We study the finite
temperature effect on different phases in Sec.\ \ref{finiteT}.
Finally, we discuss our main results, chart out experiments which
can test our theory, and conclude in Sec.\ \ref{discussion}.
Some details of our calculation and the construction of the
effective Landau theory are presented in the Appendices.

\section{Model and Method}
\label{modelmethod}

In this section, we shall present the model we use and also discuss
the details of the method used for computation.

\subsection{Model}\label{model}

We begin by defining a Rashba spin-orbit coupled two-orbital Bose-Hubbard
Hamiltonian on a square lattice in 2D:
\begin{subequations}\label{orig_H:main}
\begin{align}
  H~~ &= H_{kin} + H_U\tag{\ref{orig_H:main}}\\
  H_{kin} &= \sum\limits_{<ij>;\alpha\beta} \mathcal{R}_{\alpha\beta}(i,j)
   b^{\dagger}_{i\alpha}b_{j\beta} + H.c.\label{orig_H:K}\\
H_U~ &=
 \frac{U}{2}\sum\limits_{i;\alpha} n_{i\alpha}\left(n_{i\alpha}-1\right) +
 \lambda U \sum\limits_{i} n_{i1}n_{i2}\nonumber\\
&~~~-\sum\limits_{i;\alpha}(\mu +
\Omega\sigma_{z})n_{i\alpha}.\label{orig_H:U}
\end{align}
\end{subequations}
Here $\mathcal{R}(i,j) = -t \exp[\iota \bm{A}.(\bm{i}-\bm{j})]/2$ is
the real space hopping matrix, $\bm{A} =
(\gamma\sigma_y,-\gamma\sigma_x,0)$ is the synthetic gauge field.
$U$ is the on-site repulsion, $\lambda$ denotes the ratio between
inter-orbital and intra-orbital on-site repulsion, and $\Omega$ is
the Zeeman field which arises due to the coupling of the Raman laser
to the bosonic atom \cite{spielman2011}. This term depends on the
strength of the atom-laser coupling and can be tuned to the extent
that the spin-orbit physics does not get completely masked. In this
work, following Refs.\ \onlinecite{nandini-prl}, we shall later set
$\Omega$ to zero in order to have a clean demonstration of the
effects of spin-orbit coupling. In what follows, we also neglect
another additional on-site term $H \sim \delta \sigma_y /2$ which
depends on the detuning parameter $\delta$ of the Raman laser and
can be made small by sufficient reduction of the detuning. For the
rest of this work, we set the lattice spacing $a_0=1$.

\begin{figure}[t]
\centerline{
  \includegraphics[width=6.5cm]{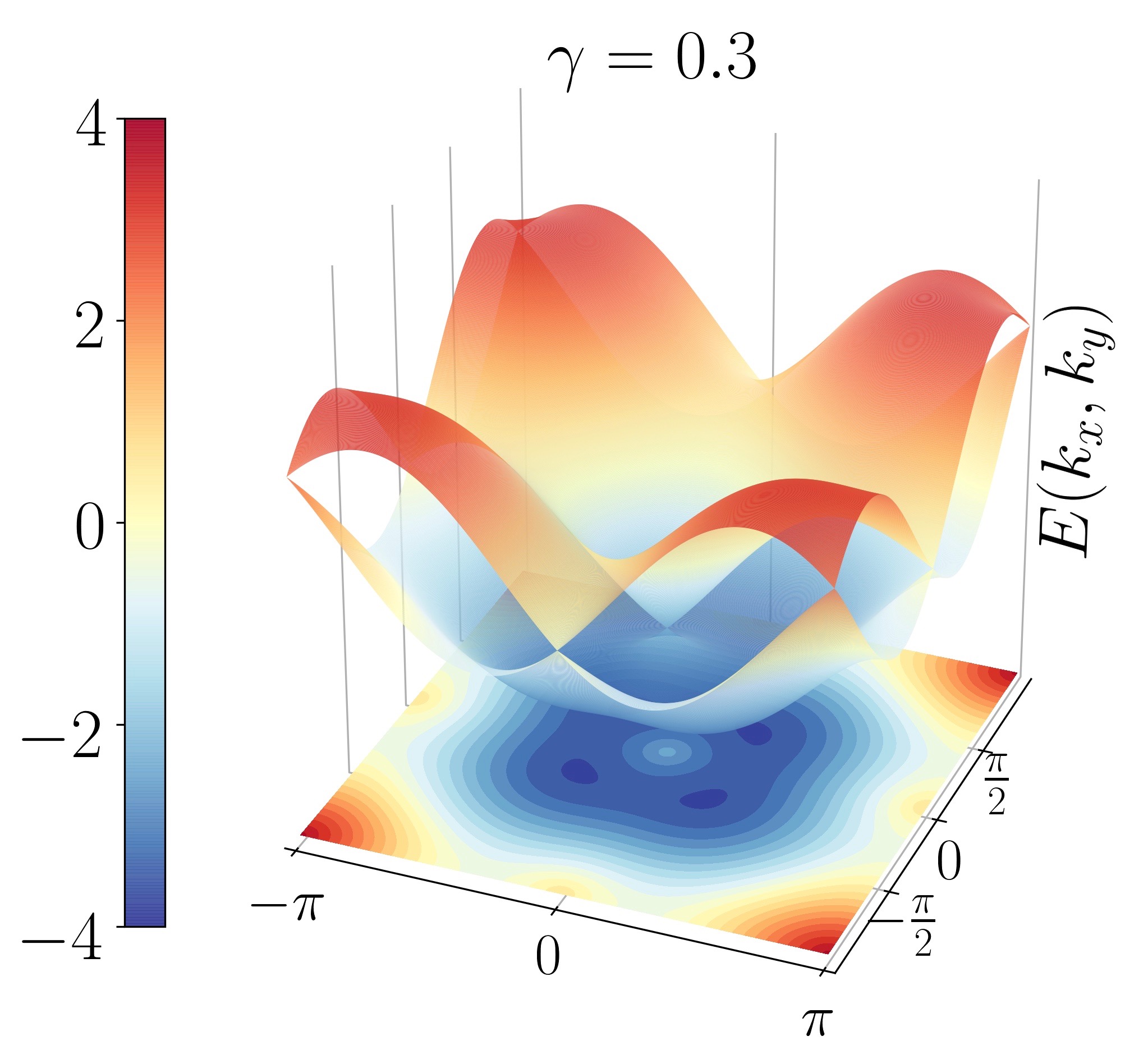}
}
\vspace{.3cm}
\centerline{
  \includegraphics[width=6.5cm,height=4.6cm]{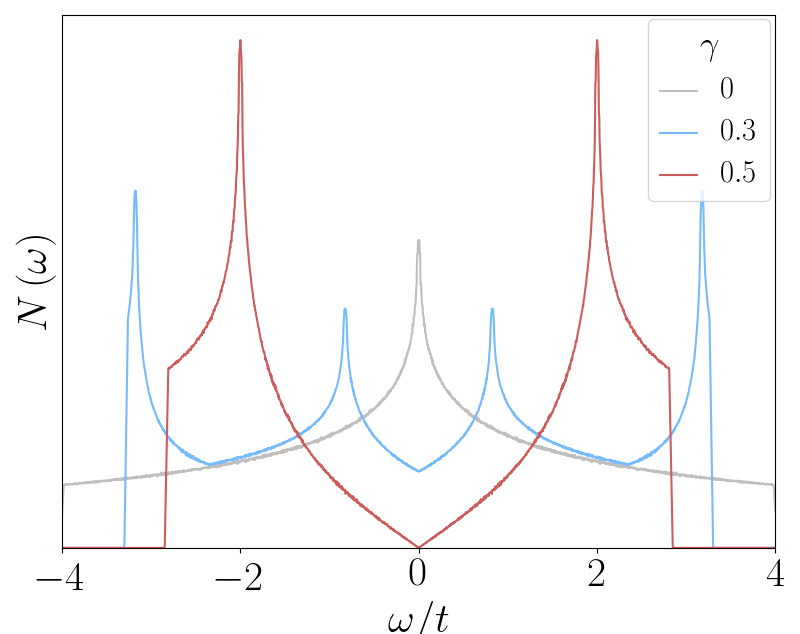}
}
  \caption{Top: The
band structure for $\gamma = 0.3$. The dispersion has a
four fold symmetry. The
minima occur at finite wavevectors, as is evident from the projection of
  the lower band onto the x-y plane. Bottom: The noninteracting
  density of states for three different values of $\gamma$. $\gamma = 0$
  has the usual tight binding form in 2D, while for finite $\gamma$ one
  observes a dip at zero along with a linearly rising behavior which is
  reminiscent of the Dirac cone present in the band structure at the
  $\Gamma$ point. All energies are in units of $t$.}\label{band}
\end{figure}

The kinetic part $H_{kin}$ can be mode separated and can be written
as
\begin{subequations}\label{Hkin:main}
\begin{align}
H_{kin} =
\sum\limits_{\bm{k}} & \irow{ b^{\dagger}_{\bm{k}1}\,\,b^{\dagger}_{\bm{k}2}}
\mathbbm{h}_{\bm{k}}\icol{b_{\bm{k}1}\\b_{\bm{k}2}}\tag{\ref{Hkin:main}}\\
\mathbbm{h}_{\bm{k}} = -2t &[\cos\gamma (\cos k_x + \cos k_y)\mathbbm{1}\nonumber\\
~~~ + \sin\gamma &(-\sin k_x\sigma_x+\sin k_y\sigma_y)]\label{Hkin:a}
\end{align}
$\mathbbm{h}_{\bm{k}}$ can be diagonalized by going to the chiral
basis. The eigenvalues and eigenvectors of $\mathbbm{h}_{\bm{k}}$
are given by
\begin{align}
E^{\pm}_{\bm{k}} &= -2t[\cos\gamma(\cos k_x+\cos k_y)\nonumber\\
&\mp (\sin\gamma\sqrt{\sin^{2}k_x+\sin^{2}k_y})]\label{Hkin:b}\\
\chi^{\pm}_{\bm{k}} &= \frac{1}{\sqrt{2}}\icol{1 \\ \pm e^{\iota
\theta_{\bm{k}}}}\label{Hkin:c}
\end{align}
with $\theta_{\bm{k}} = \tan^{-1}[\sin k_x/\sin k_y]$.
\end{subequations}
Here $+(-)$ denotes the upper(lower) bands in Fig.\ref{band}. The
band structure respects $\pi/2$ rotational symmetry of the square
lattice. Since the local interaction terms do not break this
symmetry, this degeneracy should remain intact even in the many-body
spectrum. For Rashba type spin-orbit coupling the band minima always
lie on the diagonals of the two-dimensional (2D) Brillouin zone
(BZ). The locations are at $(\pm k_0,\pm k_0)$ where $k_0$ is
determined by the strength of the SO coupling: $k_{0} =
\tan^{-1}[\tan(\gamma)/\sqrt{2}]$. The noninteracting density of
states (DOS) has been shown in Fig.\ref{band}. As the spin-orbit
coupling strength $\gamma$ is varied from $0$ to $1/2$, the DOS
develops additional van Hove singularities at finite energies, while
the singular peak at $\omega = 0$ turns into a dip with a linear
rise.

\subsection{Effective Hamiltonian}

In order to simulate the finite temperature physics of this model
we introduce auxiliary fields and implement an approximation
that maintains a positive definite stiffness for these fields.
The usual mean-field decomposition \cite{bht1} of the
kinetic term does not meet this requirement.

We start by
writing the imaginary time coherent state path integral using the
Hamiltonian above \cite{bht2}
\begin{subequations}\label{Zfull:main}
    \begin{align}
        Z &= \int\mathcal{D}\left[b^{*},b\right]e^{-\left(S^{loc}+
        S^{hop}\right)\left[b^{*},b\right]}\tag{\ref{Zfull:main}}\\
        S^{loc} &= \int_{0}^{\beta}\mathrm{d}\tau\Big [\sum\limits_{i;\alpha}b^{*}_{i\alpha}\partial_{\tau}
        b_{i\alpha}+\frac{U}{2}\sum\limits_{i;\alpha} n_{i\alpha}\left(n_{i\alpha}-1\right)\nonumber\\
        &+ \lambda U \sum\limits_{i} n_{i1}n_{i2}-\sum\limits_{i;\alpha}(\mu + \Omega\sigma_{z})n_{i\alpha}\Big ]\label{Zfull:a}\\
        S^{hop} &= \int_{0}^{\beta}\mathrm{d}\tau\left[\sum\limits_{\bm{k};\sigma
        \in \{\pm\}}\psi_{\bm{k}\sigma}^{\dagger}E_{\bm{k}}^{\sigma}\psi_{\bm{k}\sigma}\right] \,\,
         \label{Zfull:b}\\
        \psi^{+}_{\bm{k}} &= e^{\iota\theta_{\bm{k}}}b_{\bm{k}1} + b_{\bm{k}2}\label{Zfull:c}\\
        \psi^{-}_{\bm{k}} &= e^{-\iota\theta_{\bm{k}}}b_{\bm{k}1} - b_{\bm{k}2}\label{Zfull:d}
    \end{align}
\end{subequations}

\begin{figure*}
    \includegraphics[width=15cm]{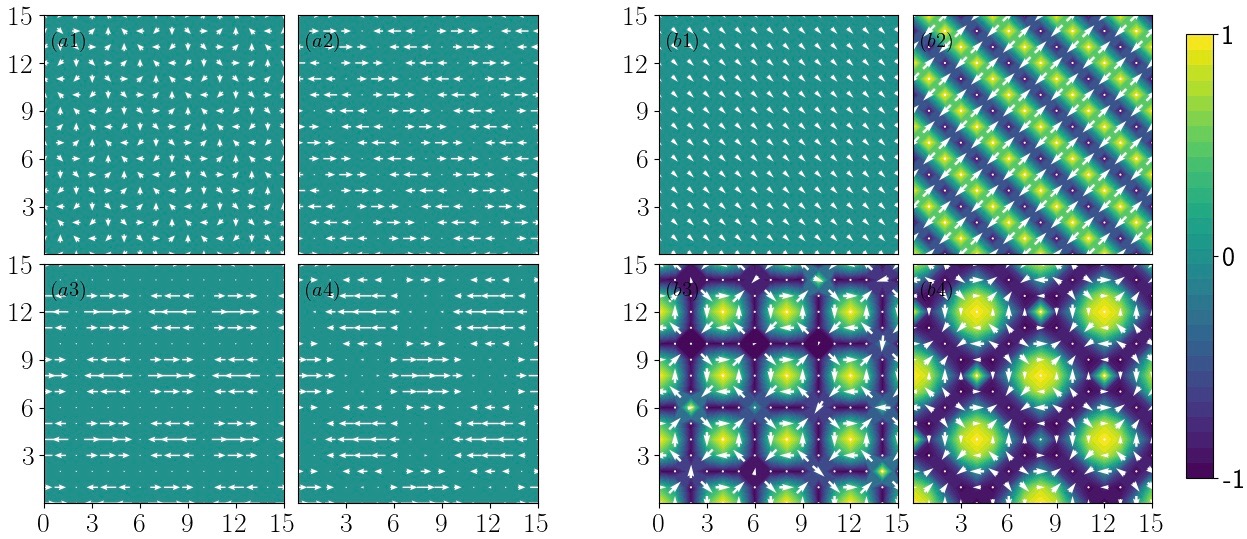}
    \caption{Left panels (a1-a4):
The variational families chosen for minimization.
The ratio $|\phi^{+}_{k_{0}}|/|\phi^{-}_{k_0}|$ has been
    plotted in color and $\left(Re\left[\phi^{-}_{k_{0}}\right],
    Im\left[\phi^{-}_{k_{0}}\right]\right)$ has been plotted using
    arrows.
Right panels (b1-b4): the  magnetic textures
corresponding to the left panels.
The $\left(\bm{m}_{x},\bm{m}_{y}\right)$
    components have been plotted using arrows, while the $\bm{m}_{z}$
    component has been plotted in color. (a1, b1) represent a typical
    single mode configuration, (a2, b2) a two mode, (a3, b3) a four
    mode, and (a4, b4) a vortex configuration. The single mode and
    the two mode configurations arise in the ground state
    but the four mode and the vortex configurations
    do not.}\label{config}
\end{figure*}

Next, we wish to implement a Hubbard-Stratonovich decomposition of
the hopping part of the action. To this end, we segregate the
negative energy part of the bands ($\tilde{E}_{\bm{k}}^{\pm}$), and
introduce an auxiliary field decomposition of the negative-band
action using two fields $\{\phi^{+}_{i,n}\}$, $\{\phi^{-}_{i,n}\}$
for each lattice point and Matsubara frequency, $(i,n)$. The effects
of positive energy part of the bands can be built back
perturbatively, and should not affect the low-energy physics
significantly\cite{joshi-thermal}. The resulting action is given by
\begin{align}
    S &= S^{loc} + \tilde{S}^{hop}\nonumber\\
    \tilde{S}^{hop} &= -\sum\limits_{k,\sigma,n}\left(\sqrt{-\tilde{E}_{k\sigma}}
    \psi^{*}_{k\sigma n}\phi_{k\sigma n} + H.c.
+ |\phi_{k\sigma n}|^{2}\right)\label{effac1}
\end{align}

Next, we note that an effective Hamiltonian can be derived from Eq.\
\ref{effac1} if we retain only the zero Matsubara frequency mode of
the auxiliary fields $\{\phi^{+}_{i,0}\}$, $\{\phi^{-}_{i,0}\}$. For
the single orbital problem this approximation reproduces the
mean-field \cite{sheshadri} ground state exactly, and captures
thermal scales which agree well with full quantum Monte-Carlo
\cite{joshi-thermal}. The effects of the finite-frequency modes can
be built back perturbatively as quantum corrections over the static
background. This has been accomplished for the single orbital
problem \cite{joshi-spectral} and such corrections are known to
leave the qualitative nature of the thermal phase and phase
transitions unchanged. For bosons coupled via spin-orbit coupling,
this turns out to be more cumbersome and we defer computation of
such corrections to a future work.

The effective Hamiltonian obtained by retaining only
$\{\phi^{+}_{i,0}\}$, $\{\phi^{-}_{i,0}\}$ fields is given by
\begin{eqnarray}
H^{\rm eff} &=& H^{eff}_{kin} + H_U \label{Heff:main} \\
H_U &=& \frac{U}{2}\sum\limits_{i;\alpha}
n_{i\alpha}\left(n_{i\alpha}-1\right) +
\lambda U \sum\limits_{i} n_{i1}n_{i2}\nonumber\\
&& -\sum\limits_{i;\alpha}(\mu + \Omega\sigma_{z})n_{i\alpha} \label{Heff:a}\\
H^{\rm eff}_{kin} &=& \sum\limits_i (\Gamma^{\dagger}_i\Psi_i +
\Psi^{\dagger}_i\Gamma_i + {|\Phi_i|}^2), \mbox{with}\\
\Gamma_i &=& \frac{1}{\sqrt{2}}
\sum\limits_j \mathcal{M}_{ji}\Phi_j \label{Heff:b}\nonumber \\
\mathcal{M}_{ji} &=&\sum\limits_{\bm{k}}e^{\iota\bm{k}\cdot(\bm{j}-\bm{i})}
\begin{pmatrix}
\sqrt{-\tilde{E}^{+}_{\bm{k}}}\, &
\sqrt{-\tilde{E}^{-}_{\bm{k}}}\\
\sqrt{-\tilde{E}^{+}_{\bm{k}}}e^{-\iota\theta_{\bm{k}}}\, &
-\sqrt{-\tilde{E}^{-}_{\bm{k}}}e^{-\iota\theta_{\bm{k}}}
    \end{pmatrix}\nonumber\label{Heff:d}
\end{eqnarray}
where $\Phi_i \equiv \icol{\phi^{+}_{i}\\\phi^{-}_{i}}$ is a local
spinor composed of zero mode of the auxiliary fields
$\{\phi^{+}_{0}\} \equiv \{\phi^{+}\}$ and $\{\phi^{-}_{0}\} \equiv
\{\phi^{-}\}$. $\Psi_i \equiv \icol{b^{1}_{i}\\b^{2}_{i}}$ is a
local spinor involving the bosons in the two orbitals.
$\mathcal{M}_{ji}$ are 2x2 matrices which couple the chiral
auxiliary fields with the orbital bosonic fields, with coefficients
picked up in the band truncation process. The information of the
spin-orbit coupling enters the effective Hamiltonian through these
coefficient matrices. Here $H_U$ is the local interaction part as in
the original Hamiltonian \ref{orig_H:U} and $\Omega$ has been set to
zero in the subsequent calculations. The details of the procedure
leading to $H_{\rm eff}$ can be found in the Appendix\ref{app-a}.

\subsection{Methods}\label{methods}

The effective Hamiltonian obtained in the last section, can be
treated using several approximation schemes. In this work, we are
going to use two such schemes. The first of these, used to obtain
zero temperature phases of the system, involves treating
$\{\Phi_i\}$ as variational parameters and subsequent minimization
 of the energy obtained from the effective Hamiltonian. In this
scheme, the energy for a configuration of $\Phi$s is obtained by
diagonalizing the boson Hamiltonian $H_{eff}[\Phi_i]$. This yields
the optimal ground state configuration of $\Phi_i$ fields. In this
work, we restrict ourselves to four families of such variational
wavefunctions given by

\begin{figure}[b]
\includegraphics[width=6.0cm,height=6cm]{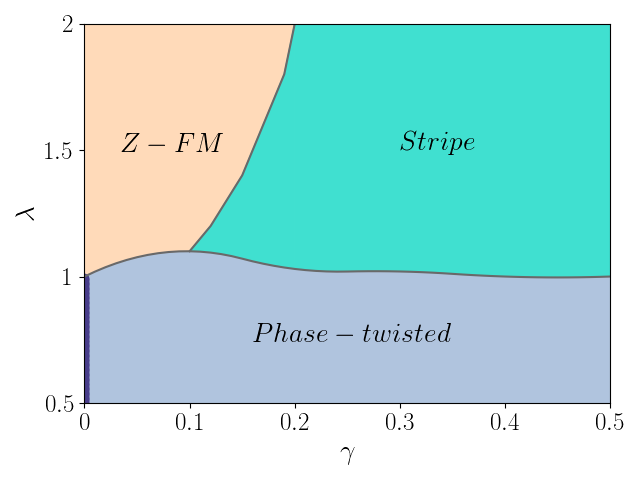}
\caption{ Classification of the ground state superfluid phases for
$U/t=10$. For $\lambda < 1$ and $\gamma = 0$ we get a homogeneous
superfluid in which $\langle b^{\dagger}_{i\alpha}\rangle$ remains
constant throughout the system. The phase-twisted superfluid has
homogeneous amplitude of $\langle b^{\dagger}_{i\alpha}\rangle$, but
its phase modulates from site to site. The Z-FM is a homogeneous
phase in which there is condensation in only one of the orbitals.
The stripe phase supports spatial modulation in both the amplitude
and the phase of $\langle b^{\dagger}_{i\alpha}\rangle$, and is
characterized by stripe-like patterns in the magnetic texture, FIG.
\ref{gsmag}. For $\gamma > 0.4$ the stripe phase shows a
$\left(\pi,\pi\right)$ order, which is the Z-AFM phase mentioned in
Ref.\ \onlinecite{nandini-prl}.} \label{magpd}
\end{figure}

\begin{enumerate}
    \item \textbf{Single mode:}
    \begin{equation*}
            \Phi_i = \icol{\phi^{+}_{k_{0}}\\
\phi^{-}_{k_{0}}}\exp(\iota \bm{k}_{0}.\bm{r}_{i})
        \end{equation*}
    \item \textbf{Two mode:}
        \begin{equation*}
\Phi_i = \icol{\phi^{+}_{k_{0}}\\
\phi^{-}_{k_{0}}}\cos(\bm{k}_{0}.\bm{r}_{i})
\end{equation*}
 \item \textbf{Four mode:}
 \begin{align*}
            \Phi_i = \icol{\phi^{+}_{k_{0}}\\ \phi^{-}_{k}}&\cos(k^{x}_{0}x_i)\cos(k^{y}_{0}y_i)
        \end{align*}
    \item \textbf{Vortex:}
        \begin{align*}
            \Phi_i = \icol{\phi^{+}_{k_{0}}\\ \phi^{-}_{k}}&\left[\cos(k^{x}_{0}x_i+
            k^{y}_{0}y_{i}) + \cos(k^{x}_{0}y_{i} - k^{y}_{0}x_i)\right]
        \end{align*}
\end{enumerate}
where ${\bf r}_i = (x_i,y_i)$ are the coordinates of site $i$. A
sketch of these variational profiles of $\Phi_i$ and the
corresponding magnetic texture of the bosons is given in Fig.\
\ref{magpd}. We note that the local Hilbert space for the bosons
needs to be restricted for the problem to be numerically tractable.
This is done by choosing a cutoff, $N_i$, in number of boson
occupation per site. In what follows, we have ensured that the
cut-off is chosen such that including more states beyond it does not
have any effect on the energy of the system, up to a desired
accuracy. The variational calculation gives us the mean field ground
state of our effective model\ref{Heff:main}.

\begin{figure}[t]
\centerline{
  \includegraphics[width=8.2cm,height=9cm]{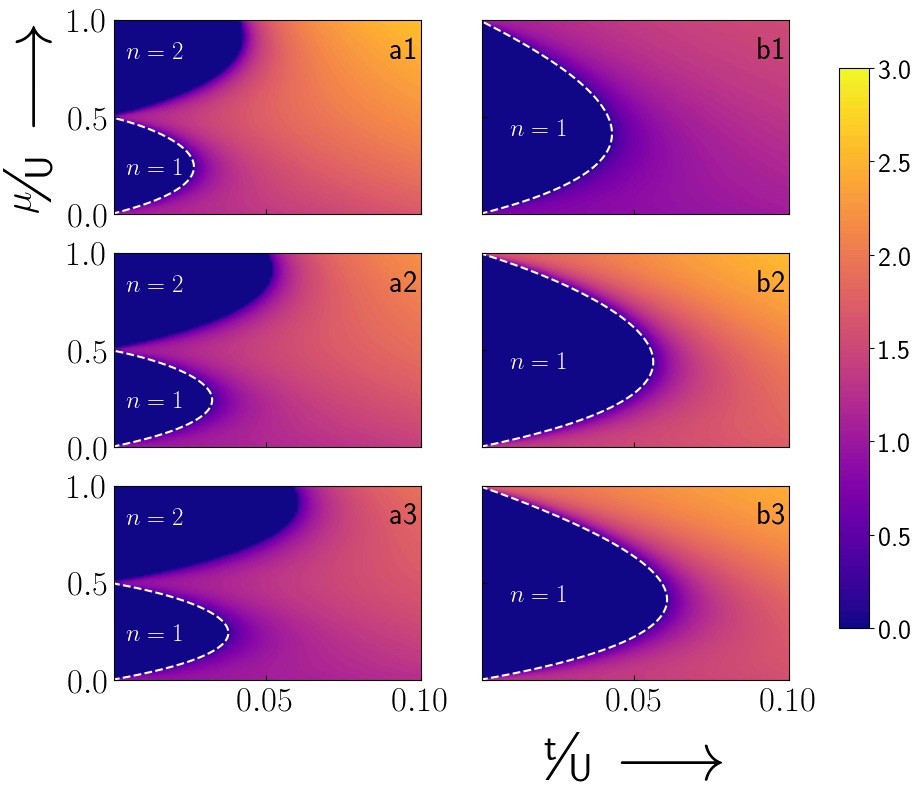}
}
\caption{Variational ground state phase diagram.
The variation of superfluid order parameter is shown in color.
Left panel a1-a3 shows the results for $\lambda$=0.5, at
$\gamma = 0, 0.3\pi$ and $0.5\pi$ respectively. The superfluid
phase in these cases is a plane wave state with homogeneous FM
order. The right panel b1-b3 shows the same plot for $\lambda$=1.5.
In this case, the superfluid phase has a two mode superposition which
leads to a stripe like magnetic texture - FIG.\ref{gsmag}.
The dashed lines demarcate the superfluid and Mott phase
boundaries  as calculated from the effective Landau
functional described
in Appendix\ref{app-b}.\label{gspd}}
\end{figure}

Having obtained the ground state configuration of the bosons, the
second method we use yields information about its thermal behavior.
To this end, we use a classical Monte-Carlo scheme by starting from
the ground state configuration and successively increasing the
temperature. The free energy for a configuration of $\{\Phi_{i}\}$s
is again obtained by diagonalizing the boson Hamiltonian $H_{eff}$
for every attempted update of the auxiliary fields. The equilibrium
$\{\Phi_{i}\}$ configurations are generated by implementing a
Metropolis based update scheme. In this scheme, at any given site
$i$, we have two complex scalar auxiliary fields, $\phi^{+}_{i}$ and
$\phi^{-}_{i}$. For each of the fields, the amplitude fluctuations
are considered to be within twice their ground state amplitude. In
contrast, arbitrary phase fluctuations of these fields are allowed.
The local hybridization $\Gamma_{i}$ depends on the $\Phi_i$
configurations on all sites, as defined in equation \ref{Heff:b}.
For a given $\{\Phi_{i}\}$ configuration the bosonic Hamiltonian is
written in Fock space after truncating the local Hilbert space
within $N_i$ particle states, as in the variational calculation. The
resulting matrix is then diagonalized exactly to obtain the free
energy for the configuration.

\subsection{Indicators}

To detect the presence of spatial order we compute the
structure factor:
\begin{equation}
    S_{\bf q} = \left<\frac{1}{zV}\sum\limits_{\bm{i},\bm{j}}Tr\left[
    \Phi^{\dagger}_{\bm{i}}\Phi_{\bm{j}}\right]
e^{\iota {\bf q} \cdot(\bm{i}-\bm{j})}\right>
\end{equation}
where $V$ is the volume of the system, $z$ is the coordination
number and $\Phi_i$s are the auxiliary fields introduced in
sec.\ref{model}.

The local magnetic texture of the
two-orbital bosons is defined by the vector,
\begin{equation}
    \bm{m}_{i} = \left<\frac{1}{Z}\sum\limits_{\mu,\nu}Tr\left[e^{-\beta H_{eff}} \,
     b^{\dagger}_{i\mu}\bm{\sigma}_{\mu\nu}b_{i\nu}\right]\right>
\end{equation}
where $Z$ is the partition function and the angular brackets denote
thermal averaging.

The momentum distribution of the bosons given by:
\begin{equation}
    n_{\bm{k}} = \frac{1}{N}\left<\frac{1}{ZV}
\sum\limits_{\bm{i},\bm{j},\mu}Tr\left[e^{-\beta H_{eff}} \,
     b^{\dagger}_{\bm{i}\mu}b_{\bm{j}\mu}\right]
e^{\iota\bm{k}\cdot(\bm{i}-\bm{j})}\right>
\end{equation}
where $N$ is the total no. of bosons, $Z$ is the partition function,
$V$ is system volume, and the angular brackets denote thermal
averaging.

\section{Variational Ground State}
\label{gs}

In this section, we shall use the variational scheme outlined
earlier to obtain the mean-field ground state phase diagram of the
bosons. In what follows, we have numerically implemented this scheme
on a $16\, \times \,16$ unit cell with $4 \le N_i \le 10$
hybridization states per site. The chosen value of $N_i$ depending
on the value of the on-site interaction $U$. For every parameter
point $N_i$ have been fixed at its optimal value, so that increasing
it does not affect the results. Unless otherwise mentioned, the
filling should be considered as fixed to one boson per site.

Due to the symmetry in the problem, we can restrict $\gamma$ to the
interval $[0, 0.5]$. Moreover, we notice that in the atomic limit,
where the problem becomes independent of $\gamma$, the level schemes
differ qualitatively if one tunes $\lambda$ across unity, as shown
in Appendix \ref{app-b} (see Fig.\ \ref{levels}). This allows us to
segregate the two parameter regimes - $\bm{\lambda < 1}$ and
$\bm{\lambda > 1}$. We present our results for a characteristic
value of $\lambda$ in each of these intervals ($\lambda= 0.5$ and
$1.5$ respectively), and expect qualitatively similar trends for
other values of $\lambda$ in the respective intervals. At each
parameter point we first classify the ground state phases using
expectation values of linear bosonic operators like $\langle
b^{\dagger}_{i\alpha}\rangle$. This allows us to demarcate the
ground state superfluid (SF) - Mott insulator (MI) phase boundary
(FIG. \ref{gspd}). The order parameter vanishes in the MI phase, as
a result, the kinetic part of the Hamiltonian has no contribution in
the energy and we recover the atomic limit. In the SF phase a
non-vanishing amplitude of $\langle b^{\dagger}_{i\alpha}\rangle$
survives throughout the system, while in the MI phase it vanishes on
all sites. We further classify the superfluid phases by using
expectation values of bosonic bilinears as defined in Eq. \ref{mag}.
This yields a classification of the superfluid phases into the
following subcategories:
\begin{itemize}
\item
\textit{Homogeneous} - where  $\langle b^{\dagger}_{i\alpha}\rangle$
and the bilinears remain constant throughout the system.
\item
\textit{Phase-twisted} - where the amplitude of
$\langle b^{\dagger}_{i\alpha}\rangle$ as well as the
bilinears remain constant throughout the system, but the phase of
$\langle b^{\dagger}_{i\alpha}\rangle$ varies from site to site.
\item
\textit{Z-FM} - in which $\langle b^{\dagger}_{i 1}\rangle$ retains a
homogeneous nonzero value, but $\langle b^{\dagger}_{i 2}\rangle$
vanishes throughout the system; $m_{z}$ remains pinned to 1, while
$m_{x}$ and $m_{y}$ vanish.
\item
\textit{Stripe} - in which both the
amplitude as well as the phase of $\langle b^{\dagger}_{i\alpha}\rangle$
vary from site to site, and the bilinears show stripe like patterns
across the system.
\end{itemize}

The effect of increasing $\gamma$ at fixed $U$ and
$\lambda$ can be understood as follows. The effective bandwidth of the
system varies with
$\gamma$ as $W \left(\gamma\right) = 4t\sqrt{2\left(1+
\cos^{2}\gamma\right)}$. Thus one requires progressively larger bare
hopping $t/U$ to compensate for the $\cos^{2}\gamma$ term in order
to stabilize the superfluid phase. Thus we expect $t_c$ to increase
with $\gamma$ for fixed $U$ and $\lambda$. This expectation is
verified in our numerics as can be seen from both panels of Fig.\
\ref{gspd}. Within the superfluid phase, the phase diagram can be
broadly classified into three separate regimes. In the first of
these, where $\bm{\lambda < 1}$ (FIG.\ref{gspd}(a)), single mode
variational profile
minimizes $H_{\rm eff}$.
For any finite value of
$\gamma$ this leads to a phase-twisted superfluid with uniform
density in both the orbitals throughout the system, while for
$\gamma = 0$ it reduces to the homogeneous superfluid phase. The fact
that any finite $\gamma$ would necessarily lead to a phase twisted
superfluid can be understood in terms of an effective Landau functional,
which has been discussed in Appendix \ref{app-b}.

\begin{figure}[t]
\includegraphics[trim={2cm 0 0 0}, width = 8cm,height=3.5cm]{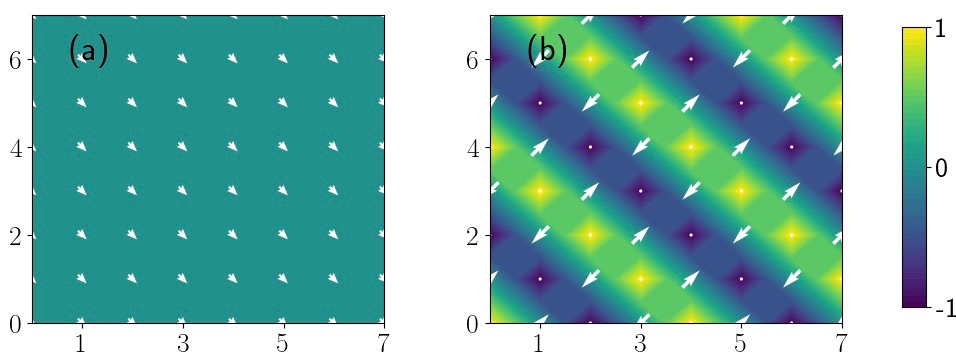}
\caption{Real space snapshot of magnetic texture in the ground state
at $\gamma$=0.3 for (a) $\lambda = 0.5$ and (b) $\lambda = 1.5$. The
$m_{z}$ component has been shown in color while the $m_{x} - m_{y}$
components have been denoted via vectors. The $\lambda$=0.5 state is
a phase-twisted superfluid with no magnetic component out of the
plane, whereas all the in-plane vectors get aligned at
$-\frac{\pi}{4}$ to the x axis. The $\lambda$=1.5 state shows a
stripe-like magnetic pattern whose pitch is controlled by the
spin-orbit coupling. } \label{gsmag}
\end{figure}

\begin{figure}[b]
 \centering
 \includegraphics[width = 8.2cm,height=5.6cm]{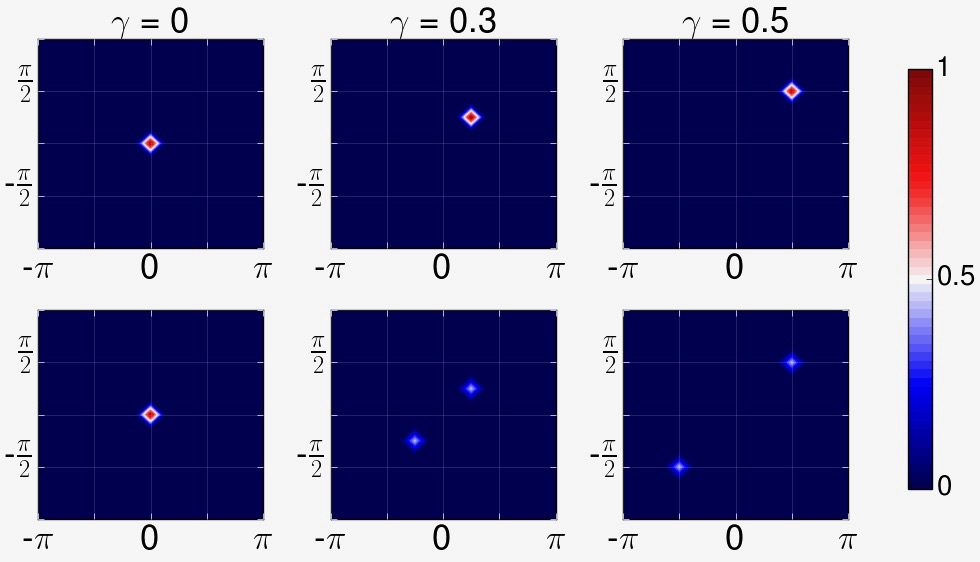}
 \caption{The ground state momentum distribution as a function of spin-orbit
     $\gamma$ has been plotted columnwise for $\gamma=0, 0.3$ and $0.5$.
     The top row shows the plots for $\lambda=0.5$ with U fixed at 10. With
      increasing $\gamma$ the condensation wave-vector moves from (0,0) to
      $\left(k_0,k_0\right)$, accompanied by slight depletion of the peak. The
      bottom row represents $\lambda=1.5$ with $u=4t$. In this case, as
      $\gamma$ is tuned from zero, the condensate splits from a single peak
      feature at (0,0) to two peaks
at $\left(-k_0,-k_0\right)$ and $\left(k_0, k_0\right)$
       with equal no. of particles at both points. The total condensate fraction,
       which has contributions from both the peaks, gets slightly depleted with
        increasing $\gamma$}\label{gsnk}
\end{figure}
In the second regime where $\bm{\lambda >
1}$ (FIG.\ \ref{gspd}(b)), for low values of $\gamma$ we get
condensation in only one of the orbitals, leading to a $z$-polarized
ferromagnetic texture as shown in . In contrast, for larger values of
$\gamma$, the two mode variational state wins over others in the
superfluid phase, leading to a stripe-like orbital order with
modulating density in each orbital. The pitch of the orbital density
wave depends of $\gamma$, and for $\gamma > 0.4$ it leads to a Z-AFM
order. The complete phase diagram in the superfluid phase as a
function of $\gamma$ and $\lambda$ is shown in Fig.\ \ref{magpd}.
The superfluid-Mott phase boundary is governed by the vanishing of
the second order coefficient of the Landau functional obtained by
tracing out the bosons in the strong coupling limit.
We discuss this procedure in detail and chart out the analytic
intuition obtained from it in Appendix\ref{app-b}.

\begin{figure}[b]
\centerline{
\includegraphics[width=8cm,height=5cm]{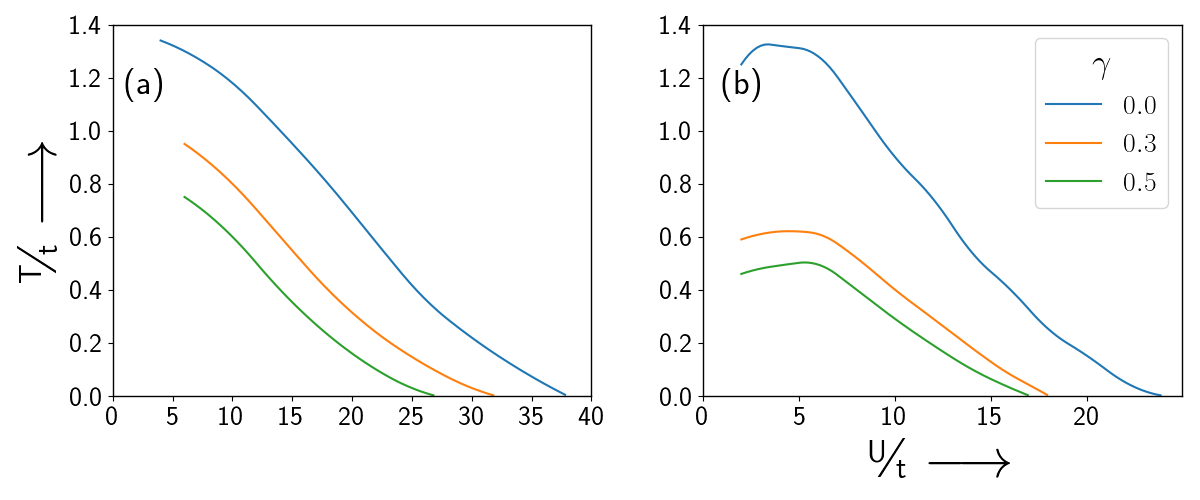}
} \caption{The thermal phase transition scales for (a) $\lambda =
0.5$ and (b) $\lambda = 1.5$. The $T_{c}(U)$ result for different
$\gamma$ are shown in color. The low temperature phase is a
superfluid with condensation at a wavevector governed by $\gamma$.
Beyond $T_{c}(U,\gamma)$ the system is a normal Bose liquid. For a
fixed $U$, $\mathrm{T}_c$ decreases with increase in $\gamma$ due to
renormalization of the bandwidth.}\label{pd}
\end{figure}
We note here that in our calculations we find that the four mode and
vortex configurations do not feature in the ground state, although
at certain parameter points their energies come very close to the
ground state energy. This is in contrast to the phase diagram
obtained in previous works \cite{nandini-prl,hofstetter} using other
techniques. This might be an artifact of band truncation in our
implementation of the mean-field approximation, although it is not
entirely clear whether other mean-field approaches can actually
capture those phases \cite{iskin}. Nevertheless, at larger values of
$\lambda\left(\gtrsim 1.5\right)$, our ground state phase diagram
matches qualitatively with that in Ref.\ \onlinecite{nandini-prl}.
In this region, we wish to highlight our finite temperature results,
since the merit of our technique is in capturing the thermal scales
nonperturbatively, which could not have been possible, to this
extent, using other techniques.

\begin{figure*}
    \includegraphics[width=\textwidth]{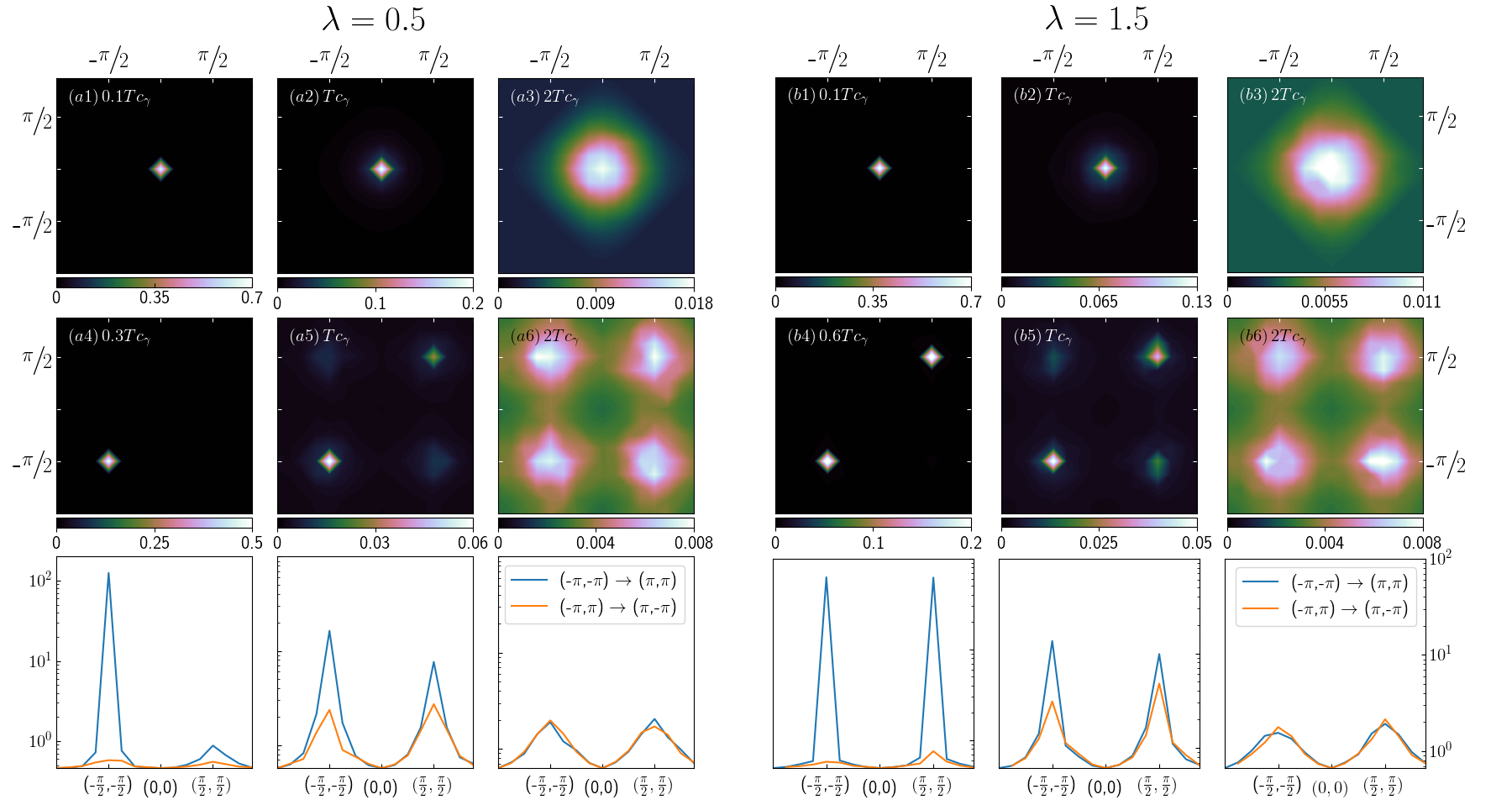}
    \caption{The thermal evolution of the momentum distribution function
($n_{\bm{k}}$) has been plotted in the left panel for $\lambda=0.5$
at $U = 14 t$, and in the right panel for $\lambda = 1.5$ for $U =
10 t$. The first two rows show evolution of the normalized
$n_{\bm{k}}$ for $\gamma = 0$ and $0.5$ respectively. The columns
show thermal broadening of the $n_{k}$ peaks as the system is heated
up from a low temperature (a1, a4) to the critical temperature
$T_{c}$ (a2, a5), and finally to a high temperature (a3, a6) where
the superfluidity has been lost. The right panel shows the same
sequence for $\lambda=1.5$. The last row shows the $n_{\bm{k}}$
projection along the two diagonals of the square BZ for $\gamma =
0.5$. For finite $\gamma$ the low temperature distribution is
sharply peaked at $(k_0,k_0)$ and $(-k_0,-k_0)$ (b4). As the
temperature reaches close to $\mathrm{T}_c$ small weights appear at
the symmetry related points $(k_0,-k_0)$ and $(-k_0,k_0)$ in the BZ
due to thermal fluctuations (b5). In the high temperature state one
can observe significant thermal broadening of the features at
relevant $k$-points (b6).}\label{nk}
\end{figure*}

Next, we study the magnetic structure of the ground state. The
magnetic texture, shown in  Fig.\ \ref{gsmag} arises from the
relative boson density modulation between the two orbitals over
different lattice sites. We find that in the ground state, for
$\lambda <1$, $m_{z i}=0$ which indicates that there is no local
population imbalance between the two orbitals throughout the lattice
as shown in Fig.\ \ref{gsmag}(a). The planar components, which
encapsulate the relative phase between the two orbitals, are also
same on all sites. In contrast, for $\lambda > 1$, the ground state,
for $\gamma=0$, has $|m_{zi}|=1$ which means that the bosons
condense in only one of the orbitals and the density in the other
orbital remains zero on all sites. Increasing $\gamma$ leads to a
diagonal stripe-like order with $|m_{zi}| <1$ indicating population
imbalance between the two orbitals. This imbalance varies in space
leading to the stripe-like order as shown in Fig.\ \ref{gsmag}(b).

At $T=0$ and in the superfluid phase, $n_{\bm{k}}$ is
sharply peaked as shown in Fig.\ \ref{gsnk}. The peak height
represents the condensate fraction, which depends on the strength of
interaction $U$ and the spin-orbit coupling $\gamma$. The condensate
gets depleted with increasing $U$ (keeping $\gamma$ and $\lambda$
fixed) leading to diminished peak height. For $\lambda < 1$, the
position of the momentum distribution peak shifts from $k=0$ to
$\left(k_{0}, k_{0}\right)$ where $k_{0}$ is given by the band
minima. This is shown in the top panel of Fig.\ \ref{gsnk}. Note
that the position of this minima is controlled by the spin-orbit
coupling. For $\lambda > 1$ the single peak at $\gamma = 0$ splits
into two peaks at $\left(\pm k_{0}, \pm k_{0}\right)$ with equal
heights as shown in the bottom panel of Fig.\ \ref{gsnk}. This
indicates that the ground state is a superposition of Bose
condensates at two distinct wavevectors. The peak heights diminish
with increasing $\gamma$, keeping U fixed. This can be attributed to
the fact that the band stiffness about the minimum decreases as the
spin-orbit strength is increased. We note that such a superposition
state may be unstable in the presence of a trap potential and we
shall not address this issue further here.

\section{Finite Temperature Results}
\label{finiteT}

In this section we chart out the finite temperature phases starting
from the variational mean-field ground states obtained in the
previous section. We use the classical Monte Carlo scheme described
in Sec.\ \ref{methods} and run the simulation on a 16$\times$16
lattice with two fluctuating fields, $\phi^{\pm i}$ and $\phi^{-}$
at each site $i$. Both the amplitude and the phase interval of the
fields are discretized  in hundred subintervals. The amplitude
interval is restricted to twice the saddle point value while full
phase fluctuation has been allowed. The real space $\{\Phi_{i}\}$
configurations are obtained by sampling over four thousand MC sweeps
for each temperature. In each these sweeps, all the sites of the
system are updated once. A total of $N_0=100$ configurations are
saved at every temperature, which are subsequently used to calculate
thermal averages of observables.\\

The finite temperature phase diagram is shown below in Fig.\ref{pd}.
The low temperature state is the variational ground state which we
have discussed at length in Sec. \ref{gs}. As we heat up the system
it gets thermally disordered and finally makes transition to a
normal state. The normal state is a Bose liquid with no long range
order, but short range spatial
correlations. The critical temperature
$T_{c}$ varies non-monotonically with $U$. As $U$ is lowered stating
from $U_{c}$, $T_{c}$ grows linearly up to quite low values of U
($\sim 2-6$ depending on $\lambda$ and $\gamma$) after which it
falls suddenly. For $\gamma = 0$ the fall is sharp and is easily
discernible in Fig.\ \ref{pd}, while for finite $\gamma$, it is
quite gradual. The low $U$ part of the phase diagram is numerically
inaccessible due to large number fluctuations in the condensate, for
which one needs to retain enormously high number of local
hybridization states. For this reason we  could access results only
up to $U/t = 2$ . With increasing $\gamma$ the $T_{c}$ scales get
suppressed at all values of $\lambda$ and U. This can again be
attributed to suppression of effective bandwidth by the spin orbit
coupling as discussed in Sec.\ \ref{gs}.

Next, we address the effect of finite temperature on the  momentum
distribution functions. The results are shown in Fig.\ \ref{nk}. The
peaks in the ground state momentum distribution show significant
thermal broadening with increasing $T$. This is best appreciated by
looking at the $\gamma = 0$ behavior (top panel in Fig.\ \ref{nk}).
The condensate fraction remains almost constant up to T =
0.1$T_{c}$, after which particles start getting excited out of the
condensate. For $T \simeq T_c$ there is significant broadening of
the peak even though the superfluid order still survives. Beyond
$T_{c}$ phase fluctuations destroy the coherence giving uniform Bose
liquid. For finite $\gamma$ one can notice thermal weights
developing in the symmetry related $k$-points when the system is
close to $T_{c}$, for both the $\lambda$ values. These weights
signify the presence of low energy states at certain $k-$ points,
which is reminiscent of the band structure symmetry. At temperatures
close to $T_{c}$ thermal fluctuations excite particles out of the
condensate to these low energy states, without destroying the
overall phase coherence in the system. As the system is heated up
further the populations in these symmetry related $k-$points tend to
homogenize at the cost of destroying superfluidity.

\begin{figure}[b]
    \includegraphics[width=8.6cm,height=3cm]{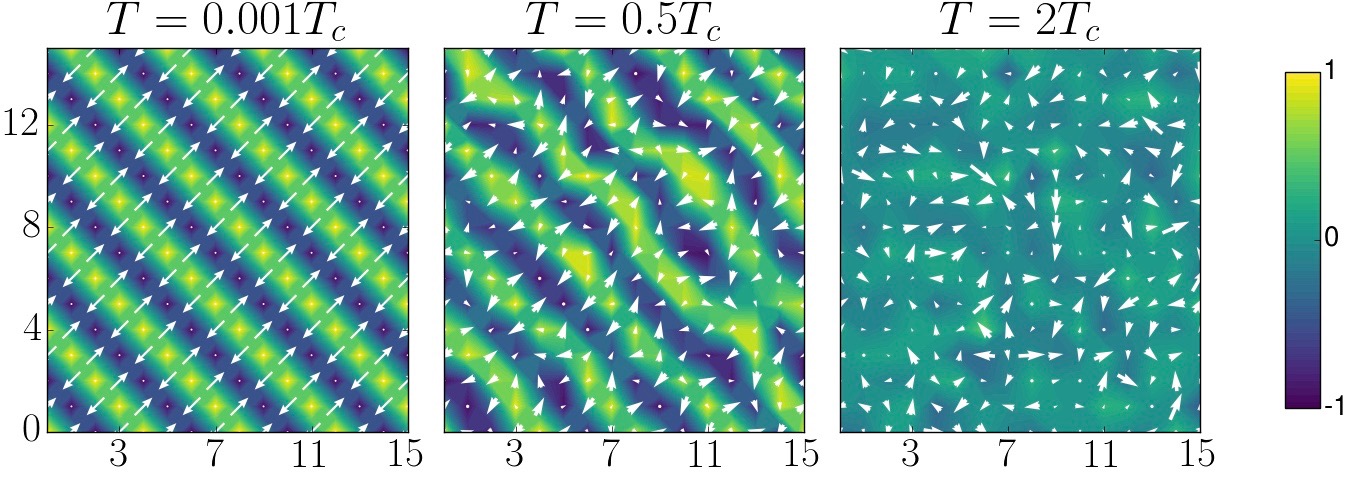}
    \caption{Spatial snapshots of $\bm{m}_{i}$ for $\lambda=1.5$ at
    $U = 10$ illustrating the temperature variation of the magnetic
    textures across the thermal transition. The orbital density wave
    survives to intermediate temperatures and vanishes for $T >> T_{c}$.
    The planar components get disordered at a lower temperature scale
    as compared to the z-component. All energies are in units of $t$.}\label{mag}
\end{figure}
Next, we consider the behavior of the magnetic texture as a function
of temperature. As the system is heated from the ground state the
magnetic textures start getting disordered. The thermal behavior of
the magnetic texture is shown in Fig.\ \ref{mag}. We observe that
for a temperature $T < T_{c}$ the planar moments become more
disordered as compared to $m_{z}$ (shown in color). This can be
attributed to the fact that the planar moments capture the gapless
phase fluctuations of the superfluid, whereas $m_z$ captures their
population difference. Finally, for $T > T_{c}$, we find that the
planar moments become completely disordered while the $z$ component
homogenizes.
\begin{figure}[t]
    \includegraphics[width=\linewidth]{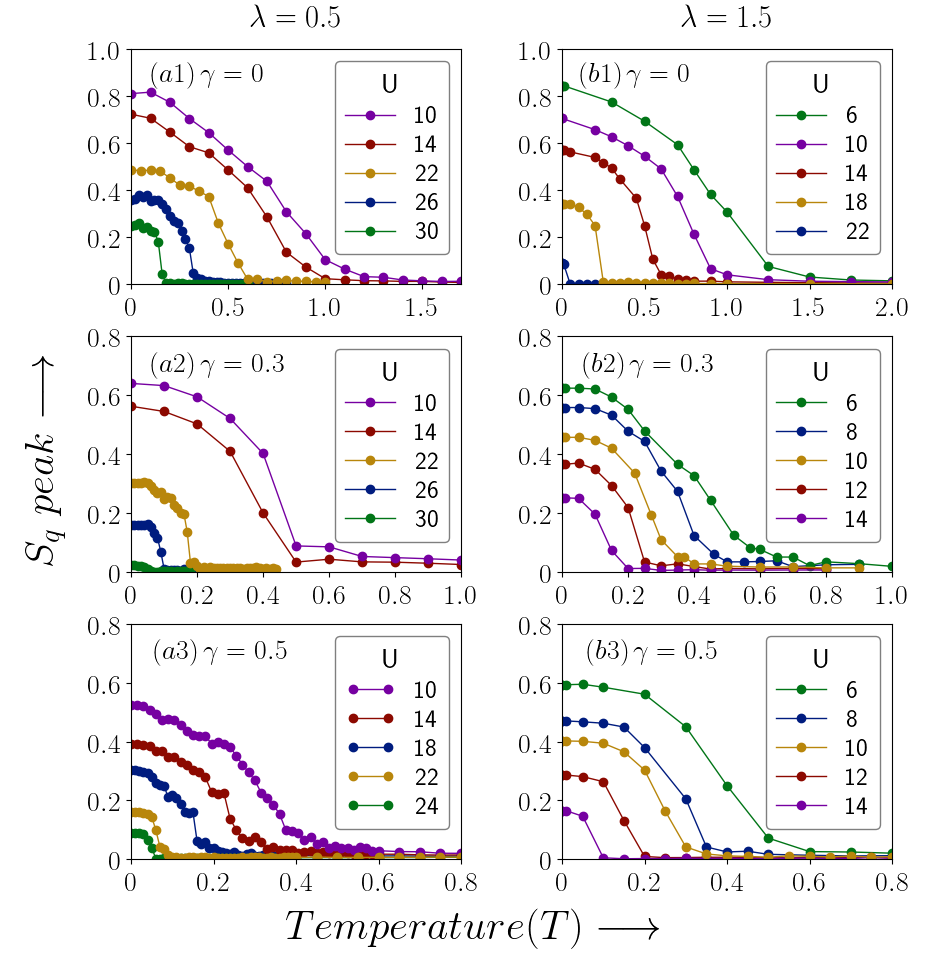}
    \caption{Thermal evolution of the structure factor peak has been
plotted for a $16 \times 16$ lattice at $\lambda=0.5$ in the first
column (a1-a3), and for $\lambda=1.5$ in the second column (b1-b3).
All energies are in units of $t$. }\label{sfac}
\end{figure}

We track the peak in the structure factor $S_{\bf q}$ with
temperature to locate the onset of long range order
 as shown in Fig.\ \ref{sfac}. We find that as the system is
heated from its ground state, the auxiliary fields start fluctuating
about their saddle point; consequently, the distribution of the
$\{\Phi_{i}\}$s broaden. At each site the two variables (per
species), i.e.\, the amplitude and the phase of the auxiliary
field $\phi_{i\sigma}$ get disordered with temperature. It is the
fluctuations of the phase degree of freedom which ultimately kill
superfluidity in the system. The transition temperature
$T_{c}\left(\lambda, \gamma, U\right)$ can be inferred from the
"knee" of the $S_{\bm{q}}$ peak vs temperature curve. Thus this
measurements allow us to locate $T_c$ which may be relevant in
realistic experiments.

\section{Discussion}
\label{discussion}

In this work we have studied the thermal phases and phase
transitions for bosons with Rashba spin-orbit coupling. Our starting
point has been a strong coupling mean-field phase of these bosons in
the SF phase near the SF-MI critical point. We find
that the result of our mean-field study lead to homogeneous,
phase-twisted, and orbital
density-wave ordered SF phases depending on the
strength of spin-orbit coupling. The phase diagram that we find
agrees qualitatively with earlier studies using more
sophisticated methods \cite{nandini-prl}. Using these ground states
as the starting point, we then perform a finite temperature Monte
Carlo study of the thermal properties of the bosons. The thermal
phase diagram for the bosons shows reduction of the critical
temperature $T_{c}$ with increasing strength of the spin orbit
coupling $\gamma$ at a fixed value of the Hubbard interaction $U$.
This can be interpreted as spin-orbit coupling introducing an
effective frustration in the system leading to reduction of order
parameter stiffness and hence $T_c$. We also obtain the thermal
broadening in the momentum distribution and the presence of
satellite peaks at the band minima which reflects the four-fold
symmetry of the Rashba term. We note that such four-fold symmetric
momentum distribution would be absent in earlier studies which
studies an effective Abelian theory involving an equal mixture of
Dresselhaus and Rashba spin-orbit terms. We find that the orbital
density waves survive to temperatures close to $T_{c}$. Finally, we
also study the magnetic textures of these bosons via computation of
the magnetization ${\bf m}_i$. In particular, we provide a clear
description of the thermal evolution of these textures and their
subsequent homogenization for $T>T_c$.

The present study neglects the quantum fluctuations of the auxiliary
fields completely. This leads to an overestimation of $U_{c}$ on one
hand, but more importantly, leads to loss of any dynamics in the
Mott phase at zero temperature. A scheme for building back the
finite frequency quantum modes already exists, and has been used to
capture quantum dynamics in the single orbital
problem\cite{joshi-thermal}. Using that method, in this problem one
hopes to recover the vortex-like magnetic textures close to the Mott
phase\cite{nandini-prl}. We leave this issue as a subject of future
study.

The simplest experimental verification of our work would be
measurement of the momentum distribution of the bosons in the SF
phase at finite temperature. We provide a detailed thermal
broadening of the momentum distribution function which could be
verified by standard experiments. In addition, we also predict that
$n_{\bf k}$ would reflect the four-fold symmetry of the Rashba
coupling term at finite temperature. This property involves peak
positions of the momentum distribution which is easily measured in
standard experiments.

{\it Conclusion:} We have studied strongly correlated two-component
bosons on a square 2D lattice in the presence of Rashba spin-orbit
coupling. We focus on the finite temperature problem and use a
recently developed auxiliary field based Monte Carlo tool, that
retains all the classical thermal fluctuations in this correlated
system, to address the thermal state. We establish, to the best of
our knowledge for the first time, the superfluid critical
temperature $T_c$ for varying intra- and inter-species repulsion and
spin orbit coupling. We study the momentum distribution and
`magnetic textures' as the temperature is increased through $T_c$
and highlight the loss of coherence and spatial order. We have
predicted experimentally verifiable signatures of the Rashba
coupling in the finite temperature superfluid.

We acknowledge use of the HPC clusters at HRI.

\appendix
\section{Derivation of effective action}\label{app-a}

The full partition function is defined in Eq.\ref{Zfull:main}.
Keeping $S^{loc}$ intact we wish to decompose the $S^{hop}$ by a
Hubbard-Stratonovich (HS) transformation. In order to implement it
we need to segregate the negative part of the bands, so that the
bosonic Gaussian integral remains well defined.
This leads to
\begin{subequations}\label{Shop:main}
    \begin{align}
        S^{hop} = S^{neg} + S^{pos}\tag{\ref{Shop:main}}
    \end{align}
    with,
    \begin{align}
        S^{neg} &= \sum\limits_{k\sigma n}\psi^{\dagger}_{k\sigma n} \tilde{E}^{\sigma}_{k}\psi_{k\sigma n}\\
        S^{pos} &= \sum\limits_{k\sigma n}\psi^{\dagger}_{k\sigma n} \left(E^{\sigma}_{k}- \tilde{E}^{\sigma}_{k}\right)\psi_{k\sigma n}
    \end{align}
    where n is the Matsubara frequency label.
\end{subequations}
    In this work, we neglect the $S^{pos}$ part and implement a HS transformation on the $S^{neg}$.
\begin{subequations}
    \begin{align}
        e^{-S^{neg}} &= e^{-\sum\limits_{k\sigma n}\psi^{\dagger}_{k\sigma n}\tilde{E}^{\sigma}_{k}\psi_{k\sigma n}}\\
        &= \prod\limits_{k\sigma n} \{\int\mathcal{D}\left[\phi^{*}_{k\sigma n},
        \phi_{k\sigma n}\right]e^{\phi^{*}_{k\sigma n}\tilde{E}_{k\sigma}^{-1}\phi_{k\sigma n}}\nonumber\\
        &\times e^{-\left(\psi^{*}_{k\sigma n}\phi_{k\sigma n}+\phi^{*}_{k\sigma n}\psi_{k\sigma n}\right)}\}\\
        \overset{\phi\rightarrow\sqrt{-\tilde{E}}\phi}{=}&
        \prod\limits_{k\sigma n} \{\int\mathcal{D}\left[\phi^{*}_{k\sigma n}
        ,\phi_{k\sigma n}\right]e^{-\phi^{*}_{k\sigma n}\phi_{k\sigma n}}\nonumber\\
        &\times e^{-\sqrt{-\tilde{E}_{k\sigma}}\left(\psi^{*}_{k\sigma n}
        \phi_{k\sigma n}+\phi^{*}_{k\sigma n}\psi_{k\sigma n}\right)}\}
    \end{align}
\end{subequations}
where $\{\phi^{+}_{n}\}$ and $\{\phi^{-}_{n}\}$ are the auxiliary
fields which couple with the respective chiral bosonic modes. This
procedure therefore leads to Eq.\ \ref{effac1} of the main text.

\begin{figure}[t]
\centering
\includegraphics[width=8cm,height=5.6cm]{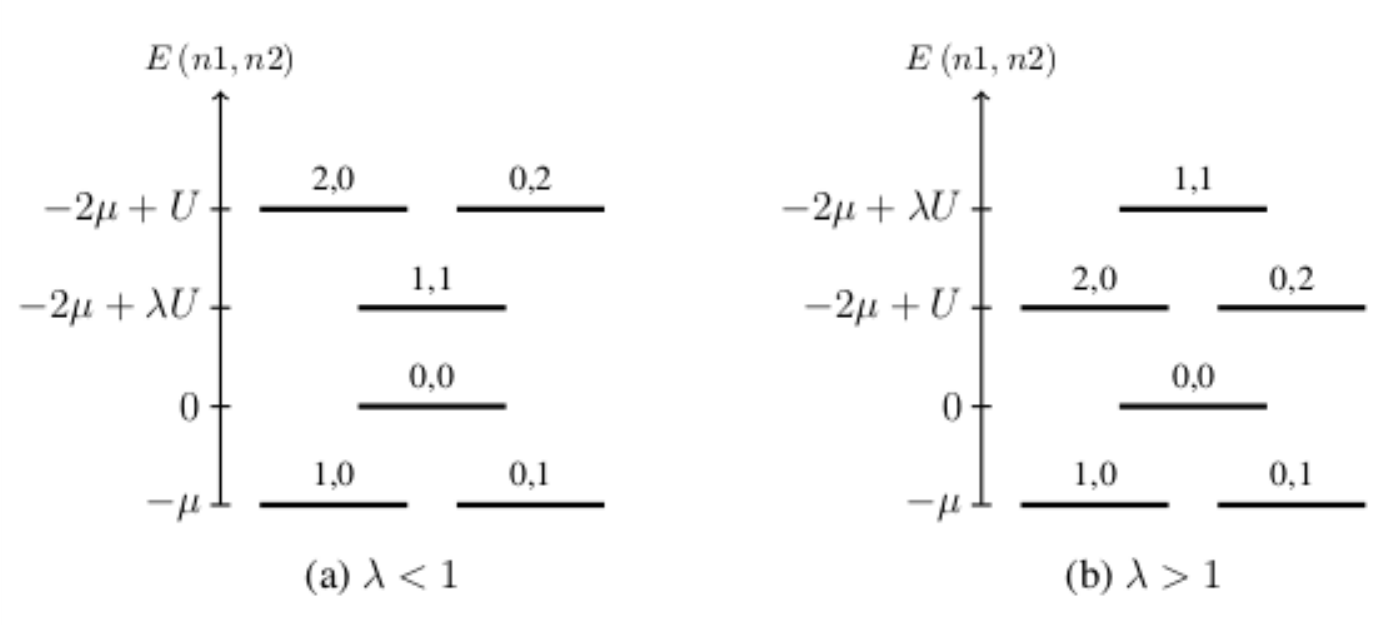}
\caption{Schematic level scheme of two-species bosons in the atomic
limit.} \label{levels}
\end{figure}

\section{Landau functional close to $U_c$}\label{app-b}

We derive an effective spin model for the bosons in the SF phase near
the SF-MI transition. To this end, note that at large $U/t$, close to
the Mott phase, the original boson fields can be integrated out to give
an effective description of the bosons in terms of the auxiliary fields.
It leads to a Landau energy functional, with coefficients depending on
the parameters of the theory. This procedure is similar in spirit to
well-known derivation of such effective spin models in the Mott phases
of the bosons\cite{demler1,issacson1}; however, here we obtain such a
model for their SF phase.

For the single orbital problem one can derive the free energy
functional by performing a cumulant expansion of the SPA functional
\cite{joshi-thermal}. In the two-orbital problem the ground state in
the atomic limit is degenerate as shown in Fig.\ \ref{levels}). Thus
one needs to use degenerate perturbation theory about the atomic
limit. The Landau energy functional after second order correction in
$\{\Gamma_{i\alpha}\}$ is given by:
\begin{align}\label{E2}
&\delta E^{(2)} = -\frac{1}{2U\tilde{\mu}}\sum\limits_{i}\biggl[
f\left(\tilde{\mu},\lambda\right)\left(|\Gamma_{i1}|^{2}+
|\Gamma_{i2}|^{2}\right)\nonumber\\
&+ \sqrt{\biggl(g\left(\tilde{\mu},\lambda\right)
\left(|\Gamma_{i1}|^{2}-|\Gamma_{i2}|^2\right)\biggr)^{2}
+\left(\frac{2\lambda\tilde{\mu}|\Gamma_{i1}\Gamma_{i2}|}{\lambda -
\tilde{\mu}}\right)^{2}}\,\,\biggr]\nonumber\\
    &+ \sum\limits_{i\sigma}|\phi_{i\sigma}|^{2} 
\end{align}
with $\tilde{\mu} \equiv \frac{\mu}{U}$, and
\begin{eqnarray}
f\left(\tilde{\mu},\lambda \right) &\equiv&
\left(\frac{1+\tilde{\mu}}{\left(1-\tilde{\mu}\right)} +
\frac{\tilde{\mu}}{\lambda-\tilde{\mu}}\right) \nonumber\\
g\left(\tilde{\mu} ,\lambda\right) &\equiv&
\left(\frac{1+\tilde{\mu}}{\left(1-\tilde{\mu}\right)}
-\frac{\tilde{\mu}}{\lambda-\tilde{\mu}}\right).
\end{eqnarray}
Notice that the square root term lifts the degeneracy of the ground
state. We now express the hybridization fields
$\{\Gamma_{i\alpha}\}$ in terms of the auxiliary fields
$\{\phi_{i\sigma}\}$ using Eq.\ \ref{Heff:b}.
\begin{align}
    |\Gamma_{\bm{i}\alpha}|^{2} =
    \sum\limits_{\bm{j}\sigma;\bm{l}\delta;\bm{k},\bm{q}}&
    \left(\left(\mathcal{M}^{\alpha\sigma}_{\bm{k}}\right)^{*}\mathcal{M}^{\alpha\delta}_{\bm{q}}\right)
    e^{\iota\left(\bm{k}-\bm{q}\right)\cdot\bm{i}+\iota\left(\bm{q}\cdot\bm{l}-
    \bm{k}\cdot\bm{j}\right)}\nonumber\\
    &\times\, |\phi_{\bm{j}\sigma}| |\phi_{\bm{l}\delta}|e^{-\iota\left(
    \theta_{\bm{j}\sigma}-\theta_{\bm{l}\delta}\right)}\label{Gamma}
\end{align}

If we choose the $\{\phi_{i\sigma}\}$ from the single mode variational
family and use the fact that the amplitude for the $\{\phi_{+}\}$
field vanishes in the ground state, then the energy functional can be
written as:
\begin{eqnarray}
\frac{\delta E^{(2)}}{V} &=&
\alpha^{(2)}\left(U,\tilde{\mu},\lambda,
\gamma\right)|\phi_{-}|^{2} \\
\alpha^{(2)}\left(U,\tilde{\mu},\lambda,\gamma\right) &\equiv&
1-\frac{|\tilde{E}^{-}_{\bm{k}_{0}\left(\gamma\right)}|}
{2U\tilde{\mu}}\left(\frac{1+\tilde{\mu}}{\left(1-\tilde{\mu}\right)}
+\frac{\tilde{\mu}\left(1+\lambda\right)}{\lambda-\tilde{\mu}}\right)
\nonumber
\end{eqnarray}
where $V$ is the volume of the system. The condensation wavevector
in the ground state is given by the $\bm{k}_{0}$ for which
$\alpha^{(2)}$ becomes maximally negative. In the expression of
$\alpha^{(2)}$ the factor in brackets remains positive definite for
the region of parameter space in which the single mode solution
dominates. Hence, the maximally negative value of $\alpha^{(2)}$
occurs at the minima of the lower band, which are given by
$\left(\pm k_{0},\pm k_{0}\right)$, with $k_{0} = \tan^{-1}\left[
\tan\gamma/\sqrt{2}\right]$. From this, we can also conclude that
the presence of an arbitrarily small $\gamma$ would lead to a
phase-twisted superfluid. At the optimal $\bm{k}_{0}$, the SF-Mott
phase boundary is determined by the zeros of $\alpha^{(2)}$. At
$\lambda = 0.5$, for which the single mode variational state
dominates, we have matched the phase boundary obtained through
numerical minimization, with that obtained from the effective Landau
theory. We find excellent agreement between the two, as is evident
in Fig.\ \ref{gspd}. A similar match was also found for $\lambda =
1.5$ where we have stripe and z-FM like order in the ground state.

\begin{figure}[t]
\centering
\includegraphics[width=8.2cm,height=4.2cm]{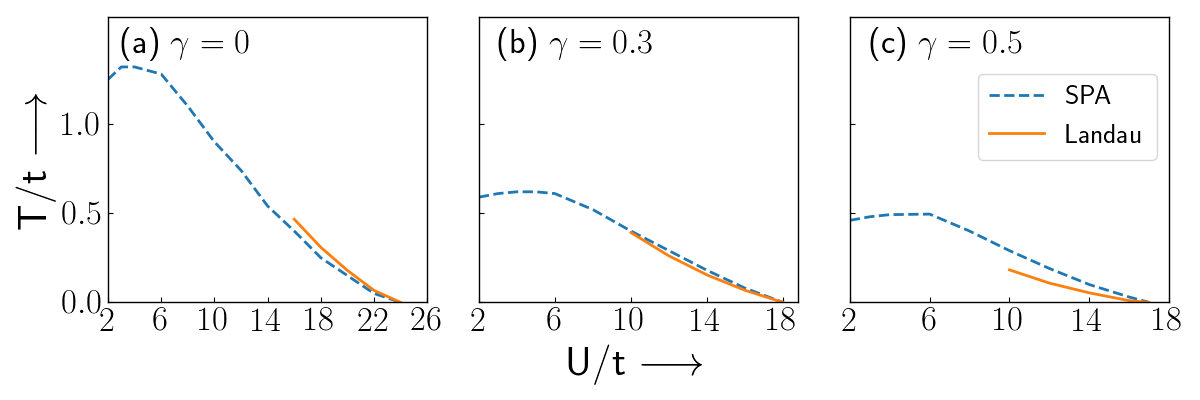}
\caption{Comparison of ordering temperatures as obtained from the SPA
based monte-carlo scheme ($T_{c}$) with that obtained from the second
order Landau functional ($T^{*}$), at $\lambda = 1.5$ for
(a) $\gamma = 0$, (b) $\gamma = 0.3$ and (c) $\gamma = 0.5$.} \label{landau}
\end{figure}

Notice that at this level we have truncated the Landau expansion to
second order. The energy functional obtained above is quadratic in
$\{|\phi_{i\sigma}|\}$, and hence the amplitudes would vanish at the
minimum. So, unless we compute the $\delta E^{(4)}$ correction, this
scheme cannot be used to optimize over the amplitudes. However, once
the optimal amplitudes are fixed from the variational calculation,
this functional may be used to anneal the phase of the auxiliary
fields, assuming that the amplitude variation with temperature is
small close to $U_{c}$. This would allow us to compare the
$T_{c}\left(U\right)$ curves of the bosonic theory with the effective
spin model. The expectation is that they would coincide at strong
coupling, as in Ref.\ \onlinecite{joshi-thermal}, allowing us to
describe the physics in terms of the low energy degrees of freedom.
For a crude estimate, one can ignore the terms
inside the square root to derive a more explicit looking functional in
terms of the phase degrees of freedom.

\begin{subequations}\label{spin:main}
\begin{align}
\tilde{E}^{(2)} &= -\frac{1}{2U}f\left(\tilde{\mu},\lambda\right)
\biggl[\sum\limits_{ij}\mathcal{A}_{ij}|\phi^{+}_{i}| |\phi^{+}_{j}|
\cos\left(\theta^{+}_{i}-\theta^{+}_{j}\right)\nonumber\\
&+ \sum\limits_{ij}\mathcal{B}_{ij}|\phi^{-}_{i}| |\phi^{-}_{j}|
\cos\left(\theta^{-}_{i}-\theta^{-}_{j}\right)\biggr]
+ \sum\limits_{i\sigma}|\phi_{i\sigma}|^{2}\tag{\ref{spin:main}}
\end{align}
with,
\begin{align}
\mathcal{A}_{ij} &\equiv \sum\limits_{\bm{k}\alpha}\left(\left(
\mathcal{M}^{\alpha +}_{\bm{k}}\right)^{*}\mathcal{M}^{\alpha +}_{\bm{k}}
\right)e^{-\iota\bm{k}\cdot\left(\bm{i}-\bm{j}\right)}\\
\mathcal{B}_{ij} &\equiv \sum\limits_{\bm{k}\alpha}\left(\left(
\mathcal{M}^{\alpha -}_{\bm{k}}\right)^{*}\mathcal{M}^{\alpha -}_{\bm{k}}
\right)e^{-\iota\bm{k}\cdot\left(\bm{i}-\bm{j}\right)}
\end{align}
\end{subequations}

The couplings $\mathcal{A}$ and $\mathcal{B}$ depend on the band
structure, and rapidly decay to zero with increasing distance. This
allows us to approximate the lattice sum by just the sum over
nearest neighbors (or the next-nearest neighbors, in case the
nearest neighbor coupling vanishes). Hence, under all these
assumptions, one can extract an effective exchange scale which would
allow us to calculate an effective ordering temperature ($T^{*}$)
for each point in our parameter space. A comparison of $T^{*}$ with
the $T_{c}$ obtained from the monte-carlo has been shown in Fig.\
\ref{landau}.  The approximation gets better at lower $\gamma$
(where neglecting the terms within the square root in Eq.\ \ref{E2}
can be easily justified) as expected. The match seems reasonably
good, given the drastic nature of approximations made for extracting
a $T^{*}$ out of the effective Landau functional.

\bibliographystyle{apsrev4-1}

\begin{thebibliography}{99}

\bibitem{rev1} I. Bloch, J. Dalibard, and W. Zwerger, Rev. Mod. Phys. {\bf 80}, 885(2008).

\bibitem{greiner1}
    M. Greiner, O. Mandel, T. Esslinger, T. W. Hansch, and I. Bloch,
    Nature (London) 415, 39 (2002); C. Orzel, A. K. Tuchman, M. L. Fenselau, M. Yasuda, and M. A.
    Kasevich, Science 291, 2386 (2001); I. B. Spielman, W. D. Phillips, and J. V. Porto, Phys. Rev. Lett. {\bf 98},
    080404 (2007).

\bibitem{expt1}
    J. Simon, W. S. Bakr, R. Ma, M. E. Tai, P. M. Preiss, and M.Greiner, Nature (London) {\bf 472}, 307 (2011);
    W. Bakr, A. Peng,E. Tai, R. Ma, J. Simon, J. Gillen, S. Foelling, L. Pollet, and M.Greiner, Science {\bf 329}, 547 (2010).

\bibitem{expt2}
    H. Bernien, S. Schwartz, A. Keesling, H. Levine, A. Om- ran, H. Pichler, S. Choi, A. S. Zibrov, M. Endres, M. Greiner, V.
    Vuletic, and M. D. Lukin, Nature (London) {\bf 551}, 579 (2017).

\bibitem{bht1} M. P. A. Fisher, P. B.Weichman, G. Grinstein, and D. S. Fisher, Phys. Rev. B {\bf 40}, 546 (1989); R. Pandit, K. Seshadri, H. R.
    Krishnamurthy, and T. V. Ramakrishnan, Europhys. Lett. {\bf 22},
    257 (1993); D. Jaksch, C. Bruder, J. I. Cirac, C. W. Gardiner, and P. Zoller,
    Phys. Rev. Lett. {\bf 81}, 3108 (1998); C.Trefzger and K. Sengupta, Phys. Rev. Lett. {\bf 106}, 095702 (2011);
    A. Dutta, C. Trefzger, and K. Sengupta, Phys. Rev. B {\bf 86}, 085140
    (2012).

\bibitem{bht2} K. Sengupta and N. Dupuis, Phys. Rev. A 71, 033629 (2005);
    J. K. Freericks, H. R. Krishnamurthy, Y. Kato, N. Kawashima,
    and N. Trivedi, ibid. {\bf 79}, 053631 (2009).

\bibitem{gauge1} Y.-J. Lin, R. L. Compton, A. R. Perry, W. D. Phillips, J. V. Porto,
    and I. B. Spielman, Phys. Rev. Lett. {\bf 102}, 130401 (2009).

\bibitem{gauge2} Y.-J. Lin, R. L. Compton,
    K. Jim\'enez-Garc\'ia, J. V. Porto, and I. B.Spielman, Nature (London) {\bf 462}, 628 (2009).

\bibitem{abeth1} Goldman N. et al., Phys. Rev. Lett., {\bf 103} 035301 (2009);
    Satija I., Dakin D. C. and Clark C. W., Phys Rev.
    lett., {\bf 97} 216401 (2006); Zhu S.-L. et al., Phys. Rev. Lett.,
    {\bf 97} 240401(2006) ; Zhai H., Umucalilar R. O. and
    Oktel M. O., Phys. Rev. Lett., {\bf 104} 145301 (2010); Umucalilar R. O. and Oktel M. O., Phys. Rev. A, 76
    (2007) 055601; Lundh E., EPL, {\bf 84} 10007 (2008);  Niemeyer M., Freericks J. K. and Monien H., Phys.
    Rev. B, {\bf 60} 2357 (1999) ; Polac T. P. and Kopec T. K.,
    Phys. Rev. A, {\bf 79} 063629 (2009).

\bibitem{abeth2} S. Sinha and K. Sengupta, Europhys. Lett. {\bf 93}, 30005 (2011);
    S. Powel, R. Barnett, R. Sensarma, and S. D. Sarma, Phys. Rev.
    Lett. {\bf 104}, 255303 (2010); K. Saha, K. Sengupta, and K. Ray, Phys.
    Rev. B {\bf 82}, 205126 (2010).

\bibitem{ketterle2016} Junru Li, Wujie Huang, Boris Shteynas, Sean
    Burchesky, Furkan Çağrı Top, Edward Su, Jeongwon Lee, Alan O.
    Jamison, and Wolfgang Ketterle Phys. Rev. Lett. {\bf 117}, 185301
    (2016).

\bibitem{ketterle2017} J.-R. Li, J. Lee, W. Huang, S. Burchesky, B. Shteynas, F. a. Top,
    A. O. Jamison, and W. Ketterle, Nature {\bf 543}, 91 (2017)

\bibitem{spielman2011} Y.-J. Lin, K. Jimenez-Garcia, and I. B. Spielman,
    Nature {\bf 471}, 83 (2011).

\bibitem{spielman-review}V. Galitski and I. B. Spielman, Nature {\bf 494}, 49 (2013).

\bibitem{iskin}A. T. Bolukbasi and M. Iskin, Phys. Rev. A {\bf 89}, 043603 (2014).

\bibitem{nandini-prl}W. S. Cole, S. Zhang, A. Paramekanti, and N. Trivedi, Phys. Rev. Lett.
{\bf 109}, 085302 (2012).

\bibitem{hofstetter} L. He, A. Ji, and W. Hofstetter, Physical Review A {\bf 92}, 023630(2015).


\bibitem{kush1} T. Grass, K. Saha, K. Sengupta, and M. Lewenstein, Phys. Rev. A
    {\bf 84}, 053632 (2011).

\bibitem{saptarshi1}  S. Mandal, K. Saha, and K. Sengupta, Phys. Rev. B {\bf 86}, 155101 (2012).

\bibitem{baym2014}G. Baym and T. Ozawa, Journal of Physics: Conference Series {\bf 529}, 012006 (2014).

\bibitem{hickey}C. Hickey and A. Paramekanti, Phys. Rev. Lett. {\bf 113}, 265302
    (2014).

\bibitem{ref51}R. A. Hart, P. M. Duarte, Tsung-Lin Yang, X. Liu, T. Paiva, E. Khatami, R. T. Scalettar, N. Trivedi,
D. A. Huse, R. G. Hulet, Nature {\bf 519}, 211-214 (2015)

\bibitem{joshi-thermal} A. Joshi and P. Majumdar, ArXiv e-prints (2017), arXiv:1711.01572 [cond-mat.str-el].

\bibitem{sheshadri}K. Sheshadri, H. R. Krishnamurthy, R. Pandit, and T. V. Ramakrishnan, EPL (Europhysics Letters) {\bf 22}, 257 (1993).

\bibitem{joshi-spectral} A. Joshi and P. Majumdar, ArXiv e-prints (2017), arXiv:1712.04433 [cond-mat.str-el]

\bibitem{demler1} E. Altman, W. Hofstetter, E. Demler, and M. D. Lukin, New J.
Phys. {\bf 5}, 113 (2003).

\bibitem{issacson1} A. Issacson, M-C Cha, K. Sengupta, and S. M. Girvin, \prb {\bf 72}, 184507 (2005)

\end{thebibliography}

\end{document}

\bibitem{pan2014}S.-C. Ji, J.-Y. Zhang, L. Zhang, Z.-D. Du, W. Zheng, Y.-J. Deng,
    H. Zhai, S. Chen, and J.-W. Pan, Nat Phys 10, 314 (2014)

\bibitem{holzmann}E. Kawasaki and M. Holzmann, “Finite temperature phases of two dimensional spin-orbit coupled bosons,” 1701.05002 [cond-
    mat.quant-gas]

To detect the presence of spatial order we compute the
boson structure factor. This is given by
\begin{equation}
    S_{\bf q} = \left<\frac{1}{zV}\sum\limits_{\bm{i},\bm{j}}Tr\left[
    \Phi^{\dagger}_{\bm{i}}\Phi_{\bm{j}}\right]
e^{\iota {\bf q} \cdot(\bm{i}-\bm{j})}\right>
\end{equation}
where $V$ is the volume of the system, $z$ is the coordination
number and $\Phi_i$s are the auxiliary fields introduced in
sec.\ref{model}.

The local magnetic texture of the
two-orbital bosons is defined by,
\begin{equation}
    \bm{m}_{i} = \left<\frac{1}{Z}\sum\limits_{\mu,\nu}Tr\left[e^{-\beta H_{eff}} \,
     b^{\dagger}_{i\mu}\bm{\sigma}_{\mu\nu}b_{i\nu}\right]\right>
\end{equation}
where $Z$ is the partition function and the angular brackets denote
thermal averaging.

The momentum distribution function for the
bosons given by
\begin{equation}
    n_{\bm{k}} = \frac{1}{N}\left<\frac{1}{ZV}
\sum\limits_{\bm{i},\bm{j},\mu}Tr\left[e^{-\beta H_{eff}} \,
     b^{\dagger}_{\bm{i}\mu}b_{\bm{j}\mu}\right]
e^{\iota\bm{k}\cdot(\bm{i}-\bm{j})}\right>
\end{equation}
where $N$ is the total no. of bosons, $Z$ is the partition function,
$V$ is system volume, and the angular brackets denote thermal
averaging.